# WAVE$^T$, A CUSTOM DEVICE ABLE TO MEASURE VISCOELASTIC PROPERTIES OF WOOD UNDER WATER-SATURATED CONDITIONS UP TO 140°C
*(WAVE$^T$ : ENVIRONMENTAL VIBRATION ANALYSER FOR WOOD)*

[1]Vincent Placet and [2]Patrick Perré *

[1] CNRS UMR 6174 FEMTO-ST Institute, Department of Applied Mechanics, 24 Ch. de l'Epitaphe. F-25000 Besançon
[2] UMR1093 Etudes et Recherche sur le Matériau Bois, INRA, ENGREF, 14, rue Girardet. F-54042 Nancy

**ABSTRACT**
This work presents an innovative experimental device conceived to characterize the time-dependent behavior of hygroscopic materials, as wood, at controlled moisture content and temperature. This device, the WAVE$^T$, permits the determination of the viscoelastic properties of samples, using harmonic tests at frequencies varying from $5.10^{-3}$ Hz to 10 Hz. Conceived to work up to 5 bar, it allows tests in dry or water-saturated conditions to be performed over the temperature range 0°C to 140°C. In spite of these severe working conditions, the careful device design, together with a rigorous data analysis, allows a rigorous determination of the storage and loss modulus and the loss factor. The results collected for several species of wood emphasize the ability of the WAVE$^T$ to underline the influence of numerous parameters, namely specie, material direction, anatomical and macromolecular structure, on the rheological properties and notably on the softening temperature. The WAVE$^T$ also establishes a efficient tool to follow the modifications of the constitutive amorphous polymers of materials submitted to hydrothermal treatments, in relation to the evolution of their viscoelastic properties.

## 1 INTRODUCTION

In comparison with quasi-static methods, harmonic tests represent a more efficient technique to characterize viscoelastic properties and glass transitions of materials. Their major advantage is to cover several decades of time independently of temperature control. In addition, this method allows the viscoelsatic behavior to be distinguished from other effects encountered with wood, such as ageing or growth stress recovery. Consequently, numerous commercial DMA apparatus using this forced non-resonance technique are available. They typically cover wide ranges of frequencies ($10^{-4}$ to 200 Hz for the most successful), and are most of the time fitted out of a hot/cold chamber. However, large discrepancies are often observed among data obtained by different dynamic mechanical analyzer, different analysis methods, or even different laboratories using identical instruments (Pournoor and Seferis, [1], Hagen et al., [2]). A number of mechanical factors, like mechanical inertia (resonance frequency), specimen geometry and size, clamping effects may influence the results of dynamic mechanical analysis. Actually, all modes of geometry (flexion, traction, shear, compression…) proposed by commercial apparatus exhibit bias due to imperfect clamping and sample geometry. To make by-pass these problems, the DMA apparatus apply correction factors to extract the viscoelastic properties of samples from the raw data. These corrective factors are fitted for isotropic material, because this type of device is intended for classical polymers. In comparison to these products, materials of vegetable origins, such as wood, present specific features, namely anisotropy and hygroscopicity. The former is problematic concerning the corrections factors and the latter is particularly important because water is recognized to act as very efficient plasticizer in hygroscopic materials.

Therefore, in the case of wood, the control of the humidity is absolutely required during dynamic mechanical analysis. Very few DMA apparatus do have humidity control, and even so, it is very difficult to keep the same moisture content of wood when the temperature changes. Notice that the moisture content variations, like drying, may cause collapse and mechano-sorptive effects. One way to avoid these problem is to proceed in a temperature-controlled water bath. The only apparatus which have the possibility to immerse the samples in fluid can only controlled the temperature up to the boiling point of the liquid, that is to say up to near 100°C for water.

Even so, an improved knowledge of the rheological properties in water-saturated conditions above 100°C is essential to develop and optimize the processing and manufacturing of wood and raw vegetable materials (drying, wood forming, thermal treatment, veneer, panel pressing, wood gluing and so on…).

For all these reasons, the wood rheology team at LERMAB developed an original experimental device, WAVE$^T$, able to perform dynamic mechanical tests of wood under water-saturated conditions and high temperature (up to 140°C with the present version). WAVE$^T$ stands for Environmental Vibration Analyzer for Wood.

* V. Placet, Department of Applied Mechanics, Femto-st Institute, University of Franche-Comté, F-25000 Besançon, vincent.placet@univ-fcomte.fr

## 2 DESCRIPTION OF THE WAVE[T]

The WAVE[T] is able to determine the evolution of the viscoelastic properties of samples with regard to temperature: 5 to 140°C, frequency: $5.10^{-3}$ to 10 Hz and stress level: 0.01 to 4 MPa. It comprises a test bench, a complete chain of vibratory analysis and a conditioning room, which allows the immersion and the temperature control of the sample during experimentation (Fig. 1).

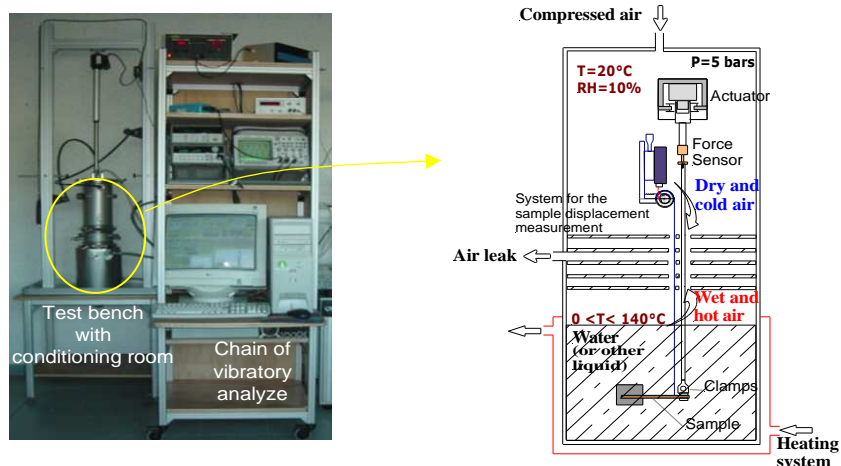

Figure 1. The WAVE[T] device.

The entire system, composed of the sample holder, the solicitation system and the sensors, is positioned in a pressurized conditioning room. This specific chamber allows the sample to be immersed in water at a temperature level reaching 140°C while maintaining mild conditions around the electronic apparatus (temperature near 20°C and relative humidity around 10%). To carry out the sample solicitation and accurately measure the force and the deflection, non-airtight communicating areas exist between the two compartments. This conditioning system, described in Fig.1, is patented (Placet and Perré, [3]).

In spite of these severe working conditions, a lot of precautions were carried out in order to ensure the relevance of the system and to allow a rigorous determination of the viscoelastic properties. The harmonic force is applied to the sample using a mini-shaker. In most commercial instruments, the force applied to the sample is deduced from the input signal to the electro-magnetic coils of the driver. In our device, a miniature load cell is used to measure the acual force. As additional advantage, a feedback loop was built with suitable PID parameters, so the force applied to the sample perfectly follows the generator signal. In this way, it is possible to apply the desired sinusoidal force at very low frequency levels: tests have been performed successfully at 1 mHz. The sole limitation is indeed the time required to get experimental data. To ensure rigorous mechanical measurements, a specific articulated clamp was designed to apply a pure vertical force without any momentum to the sample and without any vertical play. Therefore, a pure single cantilever configuration is obtained. Finally, in order to avoid any disturbance due to the clamp area, the deflection measurement is dissociated from the load system.

To keep the advantages of these experimental precautions, great care was taken to collect, treat, and analyze the raw data. The viscoelastic parameters (storage modulus E', loss modulus E'' and loss factor tan$\delta$) are determined taking into consideration of the macroscopic dimensions of the samples at 20°C, the measured force and deflection, and the phase difference. The relevance of the measurements mostly depends on how accurately the phase difference can be obtained. In particular, the method should be able to withdraw the experimental noise level. This is why an inverse method was chosen to extract this phase difference over several periods of the sinusoidal signals. Moreover, to correct or reject measurements done close to the resonance frequency, an analytical model was developed (Fig. 2.). It allows the phase change due to the inertia of all moving parts, the damping effects (water and rubber membrane of the shaker) and the frame stiffness to be dissociated from the sample properties.

All the operations are controlled by a software developed in Visual Basic, which automatically pilots the different apparatus and uses a DLL module (Dynamic Link Library) written in Fortran to identify the parameters. This application also includes the analytical model. The operator has just to key in the desired experimental protocol and to retrieve the file containing the treated data at the end of the experiment.

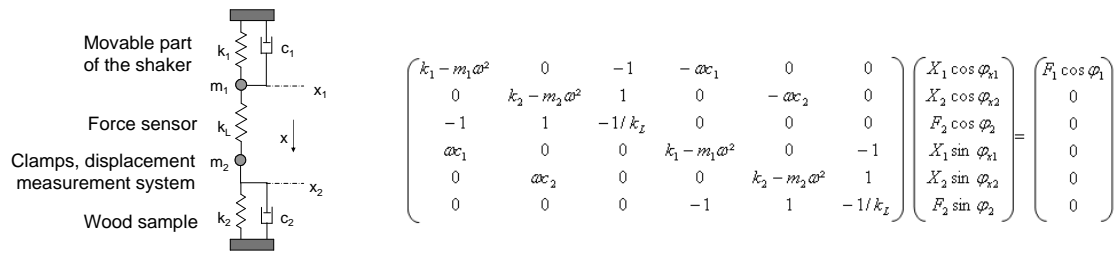

Figure 2. Diagram of the comprehensive model and its constitutive equations

## 3 SOME RESULTS

The accuracy of this apparatus has been tested using standard samples of polymer. For measurements performed on wood samples, a comparative study with a commercial DMA apparatus allowed the advantages of WAVE$^T$ to be confirmed (Placet [4]).

Several measurement campaigns have been performed on various wood species using the WAVE$^T$ apparatus (Placet and Perré, [5,6]). Results revealed a transition in the vicinity of 70 °C to 105°C depending on the wood species, the wood directions and the frequency (Fig.3a). This transition is related to the glass transition of lignins, one of the three main constitutive polymers of wood (Olsson and Salmén, [7]). In these tests, it is expected that the temperature level solely acts as a mean of activation of the viscoelastic properties. Nevertheless, other tests, performed at constant temperature over several hours, proved that the temperature also alters the material itself. Contrary to what is generally expected, this effect is significant from 100°C (Fig.3b).

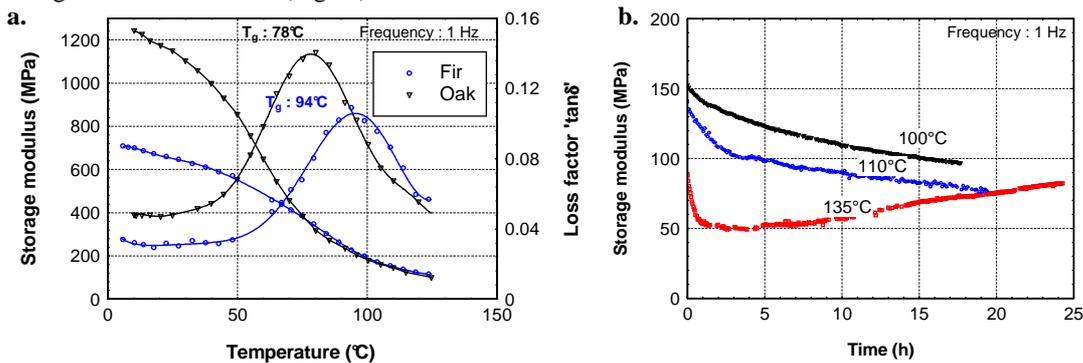

Figure 3. Evolution of the viscoelastic properties as a function of temperature (a) and time (b) for oak samples, in water saturated conditions and radial direction.

## 4 CONCLUSIONS

During the two last years, by collecting a lot of data in routine on wood samples, the WAVE$^T$ achieved a good level of confidence. The collected data and parameters could be used to predict and simulate the behaviour of green wood in drying, forming and peeling processes. The unique ability of WAVE$^T$ to assess the thermal degradation during the treatment could also be applied to define the best parameters for the heat treatment of wood in liquid water (temperature/time pathway). The increasing interest to produce ethanol from lignified biomass is certainly a good motivation for this application. In the future, an important improvement will be implemented: a controlled gas supply, based on a mixture of dry gas and saturated gas (Perré *et al.* 2007, [8]), to accurately control the relative humidity and their variations during the tests.

**REFERENCES**
1. Pournoor K. and Seferis J.C., *Polymer* **32** (3), 445-453, 1991.
2. Hagen R., Salmén L., Lavebratt H. and Stenberg, B., *Polymer Testing* **13** (2), 113-128, 1994.
3. Placet V. and Perré P., Patent. "Chambre d'essai bi-climatique". Réf: BFF 06P0432.
4. Placet V., Ph.D. report, University of Nancy I, 330 p., 2006.
5. Placet V., Passard J. and Perré P., *Holzforschung* **61**, 548-557, 2007.
6. Placet V., Passard J. and Perré P., Submitted to *Journal of Materials Science*, 2007.
7. Olsson A-M., Salmén L., Chap. 9 of *Viscoelasticity of Biomaterials*, ACS, 133-143, 1992.
8. Perré P., Houngan A.C. and Jacquin Ph., *Drying technology* **25**, 1341-1347, 2007.